\documentclass[aps,prb,showpacs,amsmath,twocolumn,amssymb,superscriptaddress,letterpaper]{revtex4}
\usepackage{mathrsfs}
\usepackage{graphicx}
\usepackage{graphics}
\usepackage{subfigure}
\usepackage{bm}
\usepackage{dcolumn}
\usepackage{amsmath,bm}
\newcommand{\nd}{{\vphantom{\dagger}}}

\usepackage{color}

\begin{document}
\title{ Detect Spinons via Spin Transport}
\author{Chui-Zhen Chen}
\affiliation{ Institute of Physics, Chinese Academy of Sciences,
Beijing 100190, China}
\author{Qing-feng Sun}
\affiliation{ Institute of Physics, Chinese Academy of Sciences,
Beijing 100190, China}
\author{Fa Wang}
\affiliation{International Center for Quantum Materials and School of Physics, Peking
University, Beijing 100871, China}
\author{X. C. Xie}
\affiliation{International Center for Quantum Materials and School of Physics, Peking
University, Beijing 100871, China}
\date{\today}

\begin{abstract}
Existence of spinons is the defining property of quantum spin liquids.
These exotic excitations have (fractionalized) spin quantum number and
no electric charge, and have been proposed to form Fermi surfaces
in the recently discovered organic spin liquid candidates.
However direct probes for them are still lacking.
In this paper we propose to experimentally identify the spinons
by measuring the spin current flowing through the spin liquid
candidate materials, which would be a direct test for the
existence of spin-carrying mobile excitations.
By the nonequilibrium Green function technique
we evaluate the spin current through the interface between a Mott
insulator and a metal under a spin bias,
and find that different kinds of Mott insulators, including
quantum spin liquids, can be distinguished
by different relations between the spin bias and spin current,
In the end we will also discuss relations to experiments and
estimate experimentally relevant parameters.
\end{abstract}

\pacs{75.10.Jm,72.25.Mk,73.23.-b, 73.40.Rw}

\maketitle

Quantum spin liquid(QSL) was first proposed by Anderson as an
alternative ground state against long range magnetic order
in frustrated magnets\cite{Anderson1973}.
In these systems competing spin exchange interactions result in a
large degeneracy of classical ground states, and quantum fluctuations
among these states destroy long range symmetry breaking order.\cite{Balents2010}
A particular kind of quantum spin liquid, the resonant valence bond(RVB) state,
has also been proposed to be the key to the high-temperature superconductivity
in the cuprate materials.\cite{Anderson1987,Lee2006}
After decades of intense research
many numerical evidences of QSL ground states have been found in
semi-realistic lattice models\cite{triangular-MSE-ED,Sheng-triangular,
 Sorella-triangular,Sheng-triangular-MSE,triangular-Hubbard-spin-ED,
kagome-ED,Sheng-kagome,White-kagome,JiangHC-square,triangular-Hubbard,Assaad},
and many artifical parent Hamiltonians for spin liquids have been
constructed\cite{Klein,Kitaev,YaoH-LeeDH}.

On the other hand, the experimental realization of spin liquids in
more than one spatial dimensions remains challenging until
several candidate materials have been discovered recently.\cite{Balents2010,Lee2008}
Two two-dimensional(2D) triangle lattice organic salts
EtMe$_3$Sb[Pd(dmit)$_2$]$_2$ and
$\kappa$-(BEDT-TTF)$_2$Cu$_2$(CN)$_3$,
the kagome lattice herbertsmithite ZnCu$_3$(OH)$_6$Cl$_2$
and a three-dimensional hyper-kagome lattice Na$_4$Ir$_3$O$_8$ are
found to be the most promising candidates of QSL.\cite{Shimizu2003,Okamoto2007,Helton2007,Itou2008}
Despite of structural distinction, they are all Mott insulators with
competing interactions and show no magnetic order down to much lower
temperature than their exchange interaction. Current measurements of
magnetic susceptibility, specific heat, thermal transport and neutron scattering
have provided vital information about the properties of these materials.\cite{M.Yamashita2008,Pratt2011,S.Yamashita2008,S.Yamashita2011}
However, there is still no definitive experiment for
the identification of quantum spin liquids.\cite{Balents2010}

One of the most significant features of quantum spin liquids is that
they have exotic excitations called spinons which are uncharged and usually
spin-1/2 mobile particles, may obey bosonic or fermionic
statistics and may have a gap or not.\cite{Balents2010,Lee2008} The fermionic
spinons may form Fermi surfaces and are generally accompanied by an
emergent gauge field.\cite{Motrunich2005,Lee2005}
This ``spinon Fermi sea'' state has received
strong support from the observations of metallic-like specific heat and thermal conductivity\cite{Yamashita2010} in
the organic candidates at low temperatures.
However these experiments do not provide a direct proof
that the mobile and possibly fermionic low energy excitations are spinons.
A more reliable proof for the spinon Fermi sea
would be the metallic-like spin transport in these Mott insulators.

\begin{figure}[htb]
\centering
\includegraphics[scale=0.85]{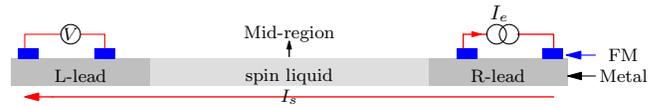}
\caption{(Color online). Schematic plot of a four-terminal device
for the detection of spin current through a spin liquid.
Four small blue bars indicate ferromagnetic electrodes.
A spin polarized current is injected into the right lead(R-lead) by a current
source($I_e$), thus creates a spin bias and drives a spin current
through the spin liquid middle region.
The spin bias induced by the spin current in the left lead(L-lead) is
then measured by a voltmeter($V$).
We will also consider other kinds of Mott insulators as the middel region
instead of spin liquids.
\label{device} }
\end{figure}

In this paper, we propose a four-terminal measurement of spin current
through a spin liquid material (Mott insulator) as the evidence for the existence of spinons.
The proposed four-terminal device consists of a spin liquid material coupled to
the left and right normal metal leads and
each lead couples to two ferromagnetic(FM) electrodes as shown in
Fig.~\ref{device}.\cite{Tombros2007}
A current source is added between the two right FM electrodes.
This is used to create a spin polarized current
flowing from one right FM electrode through the right lead to
another right FM electrode, leading to a spin-solved chemical potential
({i.e.} a spin bias $V_{R}$) in the right lead.
Then this spin bias will drive a spin current flowing from the
right lead through the spin liquid and finally into the left
lead by the spinon-electron spin-exchange interaction at the interfaces
between the spin liquid and the leads. At last, a voltmeter is connected
to the two left FM electrodes which are in contact to the left lead.
The voltmeter is used to measure the spin bias $V_{L}$ created by
the spin current in the left lead.
Here, the spin bias $V_{\alpha}$ $(\alpha=L,R)$ is defined as
the difference between the
spin-$\uparrow$ chemical potential $\mu_{\alpha \uparrow}$ and the
spin-$\downarrow$ chemical potential $\mu_{\alpha \downarrow}$ in
the $\alpha$-lead, i.e.
$V_{\alpha}\equiv\mu_{\alpha \uparrow}-\mu_{\alpha \downarrow}$.\cite{addr1,Lu2008}


In the rest of this paper,
we will first establish the general result of the spin current through the interface
between a Mott insulator in the middle region and a metallic right lead
with a spin bias (see Fig.~\ref{device}) by the nonequilibrium Green function technique.
We will then apply the general formalism to show that
different Mott insulators can be distinguished by different relations
between the spin current and the spin bias as well as temperature.
In the end we will also discuss relations to experiments and
estimate experimentally relevant parameters.

{\em The model Hamiltonian and formulation.}
In our theoretic analysis, we consider the model of a Mott insulator
(a spin liquid or a collinear antiferromagnet) coupled to two normal leads
 under a spin bias on the right lead.
The Hamiltonian of the system is given by
$H=H_0+H_M+H_I$,
where $H_0$, $H_M$, and $H_I$ are the Hamiltonians of the
leads (Metal), the middle region (Mott insulator), and the interfaces respectively.
$H_0$ and $H_I$ are assumed to be
\begin{equation}
H_{0}=\sum_{\alpha=L,R}\sum_{\boldsymbol{k},\sigma}
 \xi_{\alpha k \sigma}^\nd c_{\alpha \boldsymbol{k} \sigma }^{\dagger } c_{\alpha \boldsymbol{k} \sigma }^\nd,
\end{equation}
and
\begin{equation}
H_{I}=J_{I}\sum_{\boldsymbol{r_{0}}}\boldsymbol{S}_{R}(\boldsymbol{r_{0}})\cdot
\boldsymbol{S}_{M}(\boldsymbol{r_{0}})+J_{I}\sum_{\boldsymbol{r_{1}}}\boldsymbol{S}_{L}(\boldsymbol{r_{1}})\cdot \boldsymbol{S}_{M}(\boldsymbol{r_{1}}),
\end{equation}
where
\begin{equation*}
\boldsymbol{S}_{\alpha }(\boldsymbol {r}) =
\frac{1 }{2}
\underset{\mu ,\mu'}{\sum }\boldsymbol{\sigma }_{\mu \mu'}^\nd c_{\alpha \mu }^{\dagger }(\boldsymbol{r})c_{\alpha \mu'}^\nd(\boldsymbol {r}),
\end{equation*}
with $\boldsymbol{S}_{M}(\boldsymbol{r})$ the (dimensionless) spin operator
of the middle region at the position $\boldsymbol{r}$.
$c_{\alpha \boldsymbol{k} \sigma }^\nd (c_{\alpha \boldsymbol{k}  \sigma }^{\dagger})$ is creation (annihilation)
of the spin-$\sigma$ electron in the $\alpha$=(L,R)-lead.
$\xi_{\alpha k \sigma}=\varepsilon_{\alpha k}-\mu_{\alpha\sigma}$,
where $\varepsilon_{\alpha \boldsymbol{k}}$ is the electron dispersion relation
and $\mu_{\alpha\sigma=(\uparrow,\downarrow)}$  is the spin dependent chemical
potential in the $\alpha$-lead. The spin exchange interaction constants $J_{I}$
is determined by the interface properties of  Mott insulator and metal.\cite{Norman2009}
$\sum_{\boldsymbol{r_{0}}}$ means integral over the interface.
$H_M$ depends on the type of the Mott insulator we consider
and will be specified later.
We emphasize that there is {\em no} single electron tunneling term
in $H_{I}$ because the middle region is a Mott insulator.

Due to the spin exchange interaction in $H_I$,
the spin current can flow from the normal lead to Mott insulator and vice versa.
When the right lead is under a spin bias, the spin current $I_s$ flowing into the
middle region from the right lead is,
\begin{eqnarray}
I_{s} &=&-<\frac{d}{dt}\boldsymbol{S}_{T}^{z}> = i <[\boldsymbol{S}_{T}^{z},H(t)]>  \nonumber \\
&=&i\,J_{I}\sum_{\boldsymbol{r}_{0}}<[\boldsymbol{S}_{T}^{z},\boldsymbol{S}_{R}(\boldsymbol{r_{0}})\cdot \boldsymbol{S}_{M}(\boldsymbol{r_{0}})]> \nonumber\\
&=&J_{I}\sum_{\boldsymbol{r}_{0}} Re(\Gamma^{<}(\boldsymbol{r}_{0},t,t)),
\label{equ:2}
\end{eqnarray}
with $\boldsymbol{S}_{T}^{z}\equiv(\hbar/2)(N_{ \uparrow}-N_{\downarrow})
=(\hbar/2)\sum_{\boldsymbol{k}}(c^{\dagger}_{R\boldsymbol{k}\uparrow}c^\nd_{R\boldsymbol{k}\uparrow}
-c^{\dagger}_{R\boldsymbol{k}\downarrow}c^\nd_{R\boldsymbol{k}\downarrow})$.
Here, we have used the fact that $[\boldsymbol{S}_{T}^{z},H_{0}]=0$ and
defined $\Gamma^{<}(\boldsymbol{r}_{0},t,t) = i< S_{M}^{-}(\boldsymbol{r}_{0},t)
c_{R\uparrow}^{\dagger }(\boldsymbol{r}_{0},t^)\} c_{R \downarrow}^\nd (\boldsymbol{r}_{0},t)>$
with $S_{M}^{-}(\boldsymbol{r}_{0})=S_{M}^{x}(\boldsymbol{r}_{0})-i S_{M}^{y}(\boldsymbol{r}_{0})$.

In order to solve the Keldysh Green functions above,
we first apply equation of motion technique to solve
$\Gamma^{t}(\boldsymbol{r}_{0},\boldsymbol{k},\boldsymbol{k}',\tau,\tau^{\prime
}) = -i<T_{c}\{S_{M}^{-}(\boldsymbol{r}_{0},\tau)c_{R \boldsymbol{k} \downarrow}^\nd (\tau)c_{R \boldsymbol{k}^{\prime}\uparrow}^{\dagger }(\tau^{\prime})\}>$,\cite{Haug2008,Mahan}

By keeping the lowest order terms of $J_I$ we have
\begin{eqnarray*}
& &\Gamma^{t}(\boldsymbol{r}_{0},\boldsymbol{k},\boldsymbol{k}',\tau,\tau') \\
& = &\frac{- i J_{I}}{2N_{R}} \sum_{\boldsymbol{r}_{0}}
\int d\tau_{1} \chi^{t}(\boldsymbol{r}_{0},\boldsymbol{r}_{0}',\tau,\tau_{1}) g_{R\downarrow}^{t}(\boldsymbol{k},\tau,\tau_{1})\\
&& \times g_{R \uparrow }^{t}(\boldsymbol{k}',\tau_{1},\tau')\exp [-i(\boldsymbol{k}-\boldsymbol{k}')\cdot \boldsymbol{r}_{0}],
\end{eqnarray*}%
where $\chi^{t}(\boldsymbol{r}_{0},\boldsymbol{r}_{0}',\tau,\tau_{1})=-i<T_{c}[S_{M}^{-}(\boldsymbol{r}_{0},\tau) S_{M}^{+}(\boldsymbol{r}'_{0},\tau_{1})]>$
and $g_{R \sigma }^{t}(\boldsymbol{k}',\tau,\tau')=-i<T_{c}[c_{R \boldsymbol{k}' \sigma}^\nd(\tau)c_{R \boldsymbol{k}'\sigma}^{\dagger }(\tau')]>$
are contour-order Green functions for spin operator in middle region and free electrons in right lead, respectively.
$\Gamma^{<}(\boldsymbol{r}_{0},\boldsymbol{k},\boldsymbol{k}',t,t')= i<S_{M}^{-}(\boldsymbol{r}_{0},t)c_{R \boldsymbol{k}'\uparrow}^{\dagger }(t') c_{R \boldsymbol{k} \downarrow}^\nd(t)>$
can be obtained by analytic continuation from $\Gamma^{t}$,
and a Fourier transform then produces $\Gamma^{<}(\boldsymbol{r}_{0},t,t)$.
Plug the result into Eq.~(\ref{equ:2}) we have
\begin{eqnarray}
I_{s}&=&
 \frac{J_{I}^{2}N_{\perp}}{4N_{R}^{2}N_{M}^{2}}\underset{\boldsymbol{q},\boldsymbol{k},\boldsymbol{k}'}{\sum }
    A_{M}(\boldsymbol{q},\xi_{R k'\uparrow}\!-\!\xi_{R k\downarrow}\!+\!V )
    \delta _{\boldsymbol{q}_{\perp }+\boldsymbol{k}_{\perp} -\boldsymbol{k}'_{\perp }} \nonumber\\
& &\times\{[1\!+n_{B}(\xi_{R k'\uparrow}\!-\!\xi_{R k\downarrow}\!+\!V)]n_{F}(-\xi_{R k\downarrow})n_{F}(\xi_{R k'\uparrow}) \nonumber\\
& & -n_{B}(\xi_{R k'\uparrow}\!-\!\xi_{R k\downarrow}\!+\!V )n_{F}(\xi_{R k\downarrow})n_{F}(-\xi_{R k'\uparrow})\}
\label{equ:3}
\end{eqnarray}
where $N_{M}$ and $N_{R}$ are the number of unit cells in the middle region and right lead, respectively.
Here $N_{\perp}$ is the number of transverse mode(parallel to the interface) and $V=\mu_{R\uparrow}-\mu_{R\downarrow}$ is the spin bias in the right lead.
$n_{B}(\omega )$ and $n_{F}(\xi_{R k\sigma} )$ are the Bose and Fermi distribution functions, respectively.
$\delta_{\boldsymbol{q}_\perp+\boldsymbol{k}_\perp-\boldsymbol{k'}_\perp}$ indicates transverse(parallel to the interface)
momentum conservation.
The power spectrums $A_{M}$ are defined as
$ A_{M}(\boldsymbol{q},\omega ) = \int dt <S_{M}^{-}(-\boldsymbol{q},t) S_{M}^{+}(\boldsymbol{q},0)>\exp (i\omega t) / n_{B}(\omega)$,
where $S_{M}^{\pm}(\boldsymbol{q},t)$ is the Fourier transform of $S_{M}^{\pm}(\boldsymbol{r},t)=S_{M}^{x}(\boldsymbol{r},t)\pm S_{M}^{y}(\boldsymbol{r},t)$.

We note that the behavior of the spin current is mainly determined by
the power spectrum $A_M$ of the middle region.
Under appropriate conditions the momentum integrations over $\boldsymbol{q},\boldsymbol{k},\boldsymbol{k}'$ can be
approximately separated, and the transverse momentum conservation
factor $\delta_{\boldsymbol{q}_\perp+\boldsymbol{k}_\perp-\boldsymbol{k}'_\perp}$ will provide only a constant factor\cite{supp}.
The Fermi (Bose) function will be treated exactly at zero temperature and expanded in series
of $V/(k_{B}T)$ at finite temperature.
In the following parts of this paper we will apply Eq.~(\ref{equ:3})
and analyze several kinds of Mott insulators as the middle region,
including several spin liquids.

{\em Spin liquids.}
Spinons in spin liquids may or may not be gapped.
The gapped spin liquids will have exponentially vanishing spin transport
at low temperature and small spin bias.
We therefore restrict ourselves to the type of QSLs with
gapless fermionic spinons.
We describe such spin liquids by the following mean-field Hamiltonian
\begin{equation*}
H_{M}=\sum_{\boldsymbol{k},\sigma }\zeta_{k }f_{\boldsymbol{k}\sigma }^{\dagger }f_{\boldsymbol{k}\sigma }^\nd,
\end{equation*}
where $f$ are fermionic spinons and $\zeta_{k}=\epsilon _{k }-\mu_{s}$
with $\mu_{s}$ the spinon chemical potential.
The spinon dispersion $\epsilon_{k}$ may have a Fermi surface
(the ``spinon Fermi sea'' state\cite{Motrunich2005, Lee2005}) or
Dirac points at Fermi level\cite{Ran07}.

For illustration purpose we first consider a one-dimensional(1D) spinon Fermi sea state.
In 1D case the transverse momentum conservation factor in Eq.~(\ref{equ:3})
does not exist, and the momentum integrations can be done separately.
At $T=0$ and $\mu_{s}>>\omega$, since $\epsilon _{q}=\hbar ^{2}q^{2}/(2m_{s})$
with $m_{s}$ the spinon effective mass, the power spectrum $A_M$ divided by $N_{R}^{2}$
and integrated over momentum $q$ (the ``density of states of spin excitations'') is
\begin{equation*}
\frac{1}{N_{R}^2}\sum_{q}A_{M}(q,\omega )\propto \omega.
\end{equation*}
Replace it in Eq.\ref{equ:3}, we find spin current $I_{s}\propto V^{3}$ at $T=0$,
and $I_{s}\propto (k_{B}T)^{2}V$ at $T>0$ with $V<<k_{B}T$.

Now we come to the two dimensional spinon Fermi sea case, which is the
most relevant to the 2D organic spin liquid candidate materials.

\begin{figure}[thb]
  \centering
  \includegraphics[scale=0.7, bb = 160  300 550 520, clip=true ]{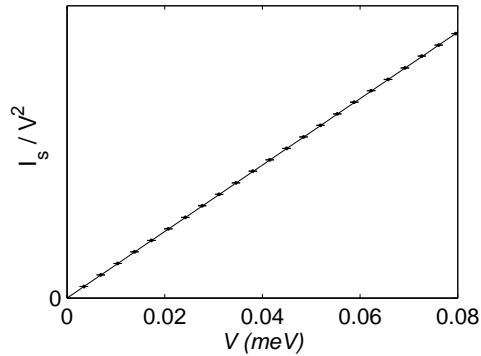}\\
  \caption{(Color online) In 2D spinon Fermi sea case, the scattered dots are the spin bias $V$ dependence of $I_{s}/V^{2}$ at zero temperature (fitted by simple line). The cut point at origin directly indicates $I_{s}\propto V^{3}$. }\label{Fig:2}
\end{figure}

\begin{figure}[thb]
  \centering
  \includegraphics[scale=0.6, bb = 20  150 450 560, clip=true ]{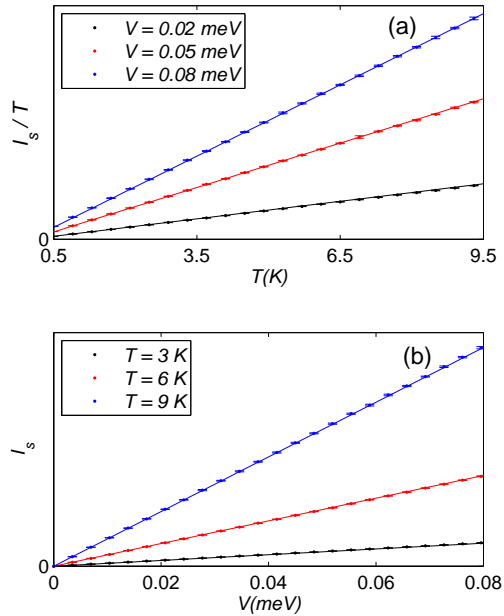}\\
  \caption{ (Color online) In 2D spinon Fermi sea case, (a) the scattered dots are temperature $T$ dependence of $I_{s}/T$ ($V$  the spin bias on the right lead).  They can be fitted by simple lines($y\propto x) $.
  (b) The spin current $I_{s}$ versus spin bias $V$ on the right lead at different temperatures $T$ (fitted by simple lines). The cut point at origin indicates $I_{s}\propto V $. (a) and (b) together indicate $I_{s}\propto T^{2}V$ at high temperature $k_{B}T>>V$.}\label{Fig:3}
\end{figure}

The spinon dispersion is $\epsilon _{q}=\hbar ^{2}q^{2}/(2m_{s})$
with $m_{s}$ the spinon effective mass,  the density of states of
spin excitations  is
\begin{eqnarray*}
\frac{1}{N_{R}^2}\sum_{\boldsymbol{q}}A_{M}(q,\omega )&=&2\pi N_{s}^{2}(E_{F}^{s})\omega \propto \omega.
\end{eqnarray*}
Thus the spin current $I_{s}\propto V^{3}$ at $T=0$ (see Fig.\ref{Fig:2}),
and  $I_{s}\propto (k_{B}T)^{2}V$ at $T>0$ with $V<<k_{B}T$ (see Fig.\ref{Fig:3}).


In the two dimensional Dirac spin liquid case\cite{Ran07},
the low energy spinon dispersion is $\epsilon _{q}=\pm
\hbar v_{F}|q-q_{F}|$, with $v_{F}$ the Fermi velocity and $q_{F}$ the Fermi vector.

At T=0, since $\sum_{\boldsymbol{q}}A_{M}(q,\omega )/N_{R}^2\propto \omega^{3}$,
the spin current $I_{s}\propto V^{5}$.
At $T>0$, $V<<k_{B}T$, expanding $I_{s}$ in the series of $V/(k_{B}T)$ directly,
we find $I_{s}\propto(k_{B}T)^{4}V$.

{\em Collinear antiferromagnetic N\'eel order.}
Antiferromagnetic(AFM) order is a common competitor for quantum spin liquids.
Let us now consider the simplest AFM ordered state,
the collinear AFM N\'eel order on a bipartite lattice,
and show that it has different spin transport behavior compared to
the previous spin liquid cases.
We describe the spin excitations by a linearized spin wave Hamiltonian,
\begin{eqnarray*}
 & & H_{M}=\sum_{\boldsymbol{k}}E_{0}(a_{\boldsymbol{k}}^{\dagger }a_{\boldsymbol{k}}^\nd
 +b_{\boldsymbol{k}}^{\dagger}b_{\boldsymbol{k}}^\nd
 +\gamma _{\boldsymbol{k}}^\nd a_{\boldsymbol{k}}^\nd b_{-\boldsymbol{k}}^\nd
 +\gamma _{\boldsymbol{k}}^\nd a_{\boldsymbol{k}}^{\dagger}b_{-\boldsymbol{k}}^{\dagger})
 \end{eqnarray*}
where $E_{0} =2ZS|J| $ and $\gamma _{\boldsymbol{k}}^\nd =\sum_{\boldsymbol{\delta} }\cos (\boldsymbol{k\cdot\delta} )/Z$
with $\delta$ sums over the nearest-neighbor lattice sites.
Here $Z$ is the coordination number, $S$ is spin quantum number and $J$ is the spin exchange constant.
$b_{\boldsymbol{k}}^\nd$ ($b_{\boldsymbol{k}}^{\dagger}$) and $a_{\boldsymbol{k}}^\nd$ ($a_{\boldsymbol{k}}^{\dagger}$),
are the annihilation (creation) of Holstein-Primakoff bosons\cite{Holstein-Primakoff}
on B-sublattice and A-sublattice, respectively.

Since spin wave dispersions is $E_{q}=4\sqrt{2}S|J|qa$ with $a$ the lattice constant
on square lattice, the density of states of spin excitations is
\begin{eqnarray*}
\frac{1}{N_{R}^2}\sum_{\boldsymbol{q}}A_{M}(q,\omega )\propto \omega ^{2}.
\end{eqnarray*}
 The spin current $I_{s}\propto V|V|^{3}$  and $I_{s}\propto(k_{B}T)^{3}V$
 at T=0, and $T>0$ with $V<<k_{B}T$, respectively.

All the cases we have discussed above are summarized in Table~\ref{tab:table1}.
\begin{center}
   \begin{table}
   \caption{
 Behavior of the spin current through the interface
 between different Mott insulators(rows) and a metallic lead
 with respect to the spin bias $V$ and temperature $T$.
}
\begin{tabular}{|c|c|c|}
  \hline
                        &$T=0$  & $k_{B}T>>V$ \\
 \hline
  1-d spinon Fermi sea & $\propto V^{3}$ & $\propto (k_{B}T)^{2}V$ \\
  \hline
  2-d spinon Fermi sea & $\propto V^{3}$ & $\propto (k_{B}T)^{2}V$ \\
  \hline
  2-d Dirac spin liquid & $\propto V^{5}$ & $\propto (k_{B}T)^{4}V$ \\
  \hline
  collinear antiferromagnet & $\propto V|V|^{3}$ & $\propto (k_{B}T)^{3}V$ \\
  \hline
 \end{tabular}

  \label{tab:table1}
 \end{table}
\end{center}

{\em Numerical estimates of the spin current.}
We use the following estimates of the parameters\cite{Kajiwara2010},
with the interface exchange coupling $J_{I}=10meV$ and
spin bias $V=1k_{B}\approx0.1meV$.
The metallic conductivity $\rho=10^{-8}\Omega\cdot m$,
lattice constant $a=0.3nm$, and Fermi level $E_{F}^{e}=1eV$.
The effective spinon mass $m_{s}\approx10m_{e}$
and Fermi level $\ E_{F}^{s}=10meV$.
The bias induced in the left lead is evaluated by
$ V_{L}\approx (2 e/\hbar) (I_{s}/N_{\perp})\ast \rho \ast l_{s} $ with $N_{\perp}$ the number of transverse mode  (see Table~\ref{tab:table2}).
\begin{table}
\caption{
 Numerical estimates of the induced spin bias in the left lead
 with different kinds of Mott insulator middle regions(rows)
 at three different temperatures $T=0K$, $1K$, and $10K$.
 Other parameters used are given in the main text.
}
 \label{tab:table2}
\begin{tabular}{|c|c|c|c|c|}
  \hline
                        & $T=0K$ & $T=1K$ & $T=10K$ \\
  \hline
  1-d spinon Fermi sea & $0.3nV$  & $10nV$ & $1000nV$ \\
  \hline
  2-d spinon Fermi sea & $0.07nV$ & $3nV$ & $300nV$ \\
  \hline
  Dirac spin liquid & $2\times10^{-7}nV$ & $5\times10^{-4}nV$ & $3nV$  \\
  \hline
\end{tabular}

\end{table}

{\em Discussions.}
Many factors ignored by our analysis may affect the results.
First we have assumed the conservation of the $z$-component of spin
in the entire system, so the spin current is well-defined.
However in the real materials spin-orbit coupling(SOC)
will generically be present\cite{Balents2010}. We hope our results can still be applied
to such systems if the linear dimensions of the sample are much smaller
than the inverse of SOC. For the same reason
we did not considered non-collinear AFM orders,
{e.g.} the 120$^\circ$ order on triangular lattice.

Secondly we have ignored the emergent $U(1)$ gauge field
in the spinon Fermi sea and Dirac spin liquid cases.
It is well-known\cite{Lee-Halperin-Read,Nave2007} that
coupling to this gauge field
can significantly change the low energy behaviors of the (spinon) Fermi sea.
However such effects have not been found in
the specific heat and thermal conductivity measurements of the organic spin liquid candidates,
the conventional Fermi liquid behaviors were observed instead\cite{M.Yamashita2008,Yamashita2010,S.Yamashita2011}.
We therefore believe our results are still valid in these materials.
The effect of $U(1)$ gauge field is an
interesting theoretical question and will be left for future studies.

Finally we have assumed a clean and free spinon or magnon system in the middle region,
without any scattering of spinons or magnons by interactions among themselves or impurites.
We think this is not a serious problem for experiments,
according to the large value of 1$\mu $m of the experimentally estimated spinon mean free path\cite{Yamashita2010}.

In summary, we propose to experimentally identify the spinons
by measuring the spin current flowing through the spin liquid
candidate materials, which would be a direct test for the
existence of spin-carrying mobile excitations.
By the nonequilibrium Green function technique
we evaluate the spin current through the interface between a Mott
insulator and a metal under a spin bias.
It is found that different kinds of Mott insulators, including
quantum spin liquids, can be distinguished
by different relations between the spin current
spin bias as well as temperature.
We hope our results can stimulate more experimental studies
of the spin liquid candidate materials
and further promote the exchange of ideas between different fields
({e.g.} spintronics and strongly correlated electrons) in condensed matter physics.

This work was financially supported by MOST of China
(2012CB921303, 2009CB929100 and 2012CB821402), NSF-China under Grants Nos.
11074174, 11121063, 91221302 and 11274364.


\end{document}